\documentclass[12pt]{article} 
\usepackage{english}

\begin{document} 

\thispagestyle{empty} 

\begin{flushright}

February 2000                                                        

\end{flushright} 

\bigskip\bigskip                                                               

\centerline{{\sl Reply to A. Patrascioiu's and E. Seiler's 
comment on our paper}}
\begin{center}
{\bf \Large{Percolation properties of the 2D Heisenberg model}}
\end{center}
\vskip 1.0truecm
\centerline{\bf
B. All\'es$^{\rm a}$, J. J. Alonso$^{\rm b}$, 
C. Criado$^{\rm b}$, M. Pepe$^{\rm c}$}
\vskip5mm
\centerline{\it $^{\rm a}$Dipartimento~di Fisica, Universit\`a di
Milano-Bicocca} 
\centerline{\it and INFN Sezione di Milano, Milano, Italy}
\vskip 2mm
\centerline{\it $^{\rm b}$Departamento de F\'{\i}sica Aplicada~I, 
Facultad de Ciencias, 29071 M\'alaga, Spain}
\vskip 2mm
\centerline{\it $^{\rm c}$Institut f\"ur Theoretische Physik,
ETH--H\"onggerberg, CH--8093 Z\"urich, Switzerland}

\bigskip\bigskip\bigskip

The most of the problems raised by the authors of the comment~\cite{0}
about Ref.~\cite{1} are based on claims which have not been 
written in~\cite{1}, for instance almost
all the introduction and the point (1) in~\cite{0} are based on such
non--existent claims.

 Instead in Ref.~\cite{1} we avoid
to make claims not based on well--founded results. For instance in the
abstract we write ``... This result indicates how the model can avoid
a previously conjectured Kosterlitz--Thouless [{\sl KT}] 
phase transition...'' and
in the conclusive part we notice that ``Our results exclude this
massless phase for $T>0.5$''. Therefore it seems to us that the opening
sentence in the Comment~\cite{0} 
``In a recent letter All\'es et al. claim to show
that the two dimensional classical Heisenberg model does not have a
massless phase.'' is strongly inadequate.

\vskip 5mm

As for the points that appear in the Comment:

\vskip 3mm

\begin{itemize}
\item{(1)} 
The purpose of the paper~\cite{1} is to fill a gap in the
research about the critical properties of the Heisenberg model. This
gap is the following one: in Ref.~\cite{2} a scenario was proposed
where the 2D Heisenberg model should undergo a KT phase
transition at a finite temperature. This scenario is based mainly on
three hypotheses, the third one (which states the 
non--percolation of the ${\cal S}$--type or equatorial clusters) 
being left in~\cite{2} without a plausible justification. To back up that
hypothesis a numerical test was cited in~\cite{2} but the details of the
numerics (temperature, size of the lattice, etc.) and several data
concerning the percolation properties of the system, were
completely skipped. The only quoted result was (see beginning of
section 4 in~\cite{2}) ``We also tested numerically for $\epsilon=1/3$,... 
There is no indication of percolation...''. On the contrary, such
interesting results about the critical properties should be put forward
with a thorough description of the hypotheses involved. Moreover,
one would like to understand how was possible to use the small value
of epsilon mentioned in Ref.~\cite{2}, because that value implies a
really tiny temperature $T$ and consequently it requires a huge
lattice size. If ``Everybody agrees
that at $\beta=2.0$ the standard action model has a finite correlation
length'', see~\cite{0}, also {\it everybody} 
would like to know details about the numerics and the computer used 
to simulate the model at such a small temperature. 

\vskip 3mm

\item{(2)} There is a statement in~\cite{1} which is repeated several 
times: all results are valid for any versor $\vec n$ 
of the internal symmetry space $O(3)$. In particular, a percolating
equatorial cluster is found for every $\vec n$.
Under these conditions, we do not see how the percolation of the equatorial
cluster may lead to a breaking of the $O(3)$ symmetry. 

 On the other hand, the fractal properties of a cluster 
are very sensitive to the choice of parameters. By varying $\epsilon$ 
around the value $\epsilon=1$ (for $T=0.5$), one can make the
data for $\langle M_{\cal S}\rangle/L^2$ in Table~1 of~\cite{1} to
change rather dramatically. It is important (even in the case of a
high temperature regime, like $T=0.5$) to study this dependence. It is
sensible to expect that the fractal properties of the cluster show up
at the threshold of percolation. Again in~\cite{1} we 
do not claim that the cluster is a fractal, but just
write ``... [{\sl the equatorial clusters}] present a high degree of
roughness recalling a fractal 
structure''. To state any firmer claim, a deep analysis of the errors
and better statistics in Table~1 should be performed. 
All these problems are currently investigated.

\vskip 3mm

\item{(3)} It is true that not all flimsy clusters can avoid a KT transition.  
However this trivial truth proves nothing. Other kinds of lattices can hold
versions of the $XY$ model with no transition (see for instance~\cite{3}). 

On the other hand, the statement ``... there should be no doubt that
on such a lattice [{\sl square holes of side length $L$}] the $O(2)$ model
has a KT phase transition for any finite $L$'' is surprising. In
Ref.~\cite{4} it is shown that for any finite $L$ the KT transition is
still present but it approaches $T=0$ as $L$ becomes larger. The idea
of a fractal as the limit of some kind of cluster should not be
forgotten.

\vskip 3mm

\item{(4)} We agree with one of the sentences of this point: ``It would
be interesting to verify this [{\sl the existence of a KT transition
for $XY$ models on a fractal lattice}]''. Yet we do not see the
relevance of such an obvious claim. 

We disagree however with the authors of~\cite{0} when they say ``our
argument does not depend on the existence of such a transition on that
particular percolating cluster''. Instead, after the conclusions
of Ref.~\cite{1}, we think that the {\it non--rigorous}
proof proposed in~\cite{2} for the case 
when the equatorial cluster does percolate,
heavily lies on whether or not such a transition is realized.

\end{itemize}
\vskip 5mm

\centerline{\bf ------------------------}


\begin{thebibliography}{99}
\bibitem{0} A. Patrascioiu and E. Seiler, unpublished report
hep--lat/9912014 (v1).

\bibitem{1} B. All\'es, J.J. Alonso, C. Criado and M. Pepe, Phys. Rev. Lett. 83
(1999) 3669.

\bibitem{2} A. Patrascioiu and E. Seiler, Nucl. Phys. B (Proc. Suppl.) 30
(1993) 184.

\bibitem{3} Yu.E. Lozovik and L.M. Pomirchy, Solid State Comm. 89 (1994) 145.

\bibitem{4} P. Minnhagen and H. Weber, Physica B152 (1988) 50.
\end{thebibliography}
\end{document}